\documentclass[a4paper,10pt]{article}
\usepackage{graphicx}
\usepackage{amssymb}
\usepackage{bm}
\begin{document}

\title{\textbf{Noise improvement of SNR gain in parallel array of bistable dynamic
systems \\ by array stochastic resonance}}

\author{Fabing Duan \footnote{fabing1974@yahoo.com.cn}
\\
\small Institute of Complexity Science, Qingdao University,
Qingdao 266071, People's Republic of China \\ Fran\c{c}ois
Chapeau-Blondeau \footnote{chapeau@univ-angers.fr} \\ \small
Laboratoire d'Ing\'{e}nierie des Syst\`{e}mes Automatis\'{e}s
(LISA), Universit\'{e} d'Angers, \\ \small 62 avenue Notre Dame du
Lac, 49000 Angers, France
\\
Derek Abbott \footnote{dabbott@eleceng.adelaide.edu.au}
\\ \small Centre
for Biomedical Engineering (CBME) and School of Electrical \&
Electronic Engineering, \\ \small The University of Adelaide, SA
5005, Australia} \maketitle

\begin{abstract}
We report the regions where a signal-to-noise ratio (SNR) gain
exceeding unity exists in a parallel uncoupled array of identical
bistable systems, for both subthreshold and suprathreshold
sinusoids buried in broadband Gaussian white input noise. Due to
independent noise in each element of the parallel array, the SNR
gain of the collective array response approaches its local maximum
exhibiting a stochastic resonant behavior. Moreover, the local
maximum SNR gain, at a non-zero optimal array noise intensity,
increases as the array size rises. This leads to the conclusion of
the global maximum SNR gain being obtained by an infinite array.
We suggest that the performance of infinite arrays can be closely
approached by an array of \textit{two} bistable oscillators
operating in different noisy conditions, which indicates a simple
but effective realization of arrays for improving the SNR gain.
For a given input SNR, the optimization of maximum SNR gains is
touched upon in infinite arrays by tuning both array noise levels
and an array parameter. The nonlinear collective phenomenon of SNR
gain amplification in parallel uncoupled dynamical arrays,
i.e.~array stochastic resonance, together with the possibility of
the SNR gain exceeding unity, represent a promising application in
array signal processing.
\end{abstract}

\section{\label{sec:I} Introduction}
The past decade has seen a growing interest in the research of
stochastic resonance (SR) phenomena in interdisciplinary fields,
involving physics, biology, neuroscience, and information
processing. Conventional SR has usually been defined in terms of a
metric such as the output signal-to-noise ratio (SNR) being a
non-monotonic function of the background noise intensity, in a
nonlinear (static or dynamic) system driven by a subthreshold
periodic input \cite{Gammaitoni1998}. For more general inputs,
such as non-stationary, stochastic, and broadband signals,
adequate SR quantifiers are information-theoretic measures
\cite{Gammaitoni1998,Peter2000}. Furthermore, aperiodic SR
represents a new form of SR dealing with aperiodic inputs
\cite{Collins1995}. The coupled array of dynamic elements
\cite{Inchiosa1996,Inchiosa1995,Lindner,Casado2006} and spatially
extended systems \cite{Jung1995} have been investigated not only
for optimal noise intensity but also for optimal coupling
strength, leading to the global nonlinear effect of spatiotemporal
SR \cite{Jung1995}. By contrast, the parallel uncoupled array of
nonlinear systems gives rise to the significant feature that the
overall response of the system depends on both subthreshold and
suprathreshold inputs
\cite{Collins&Chow,Chialvo,Moss1995,Stocks2000,Stocks2001a,Stocks2001b,Stocks2001c,Stocks2002,Mcdonnell2002,Mcdonnell2006,Rousseau}.
In this way, a novel form of SR, termed suprathreshold SR
\cite{Stocks2000}, attracted much attention in the area of
noise-induced information transmissions, where the input signals
are suprathreshold for the threshold of static systems or the
potential barrier of dynamic systems
\cite{Stocks2000,Stocks2001a,Stocks2001b,Stocks2001c,Stocks2002,Mcdonnell2002,Mcdonnell2006,Rousseau}.
In addition, for a single bistable system, residual SR (or
aperiodic SR) effects are observed in the presence of slightly
suprathreshold periodic (or aperiodic) inputs
\cite{Apostolico,Duan2004}.

So far, the measure most frequently employed for conventional
(periodic) SR is the SNR \cite{Gammaitoni1998,Peter2000,Debnath}.
The SNR gain defined as the ratio of the output SNR over the input
SNR, also attracts much interest in exploring situations where it
can exceed unity
\cite{Loerincz,Chapeau,Chapeau1999,Chapeau1997,Rousseau,Chapeau2004,Chapeau2006,Peter2000,Gingl,Gingl2001,Casado-Pascual,Casado-Pascual2003,Casado-Pascual2003b,Chizhevsky}.
Within the regime of validity of linear response theory, it has
been repeatedly pointed out that the gain cannot exceed unity for
a nonlinear system driven by a sinusoidal signal and Gaussian
white noise \cite{Jung1991,Dykman1998,Dykman1995,Neiman,Gailey}.
However, beyond the regime where linear response theory applies,
it has been demonstrated that the gain can indeed exceed unity in
non-dynamical systems, such as a level-crossing detector
\cite{Loerincz}, a static two-threshold nonlinearity
\cite{Chapeau,Chapeau1999,Chapeau1997}, and parallel arrays of
threshold comparators or sensors
\cite{Rousseau,Chapeau2004,Chapeau2006}, and also in dynamical
systems, for instance, a single bistable oscillator
\cite{Peter2000,Gingl,Gingl2001,Casado-Pascual,Casado-Pascual2003,Casado-Pascual2003b,Chizhevsky},
a non-hysteretic rf superconducting quantum interference device
(SQUID) loop \cite{Inchiosa1998}, and a global coupled network
\cite{Inchiosa1995}.

\begin{figure}
\begin{center}
  \includegraphics[scale=0.45]{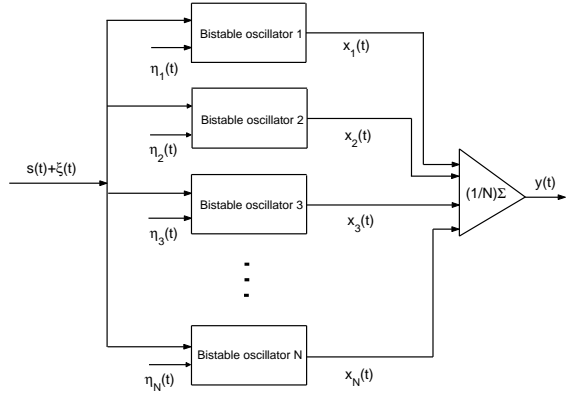}
  \caption{A parallel array of $N$ archetypal over-damped
bistable oscillators. Each oscillator is subject to the same noisy
signal but independent array noise. In this paper, we call
$\xi(t)$ the input noise and $\eta_i(t)$ the array noise.
\label{fig:one}} \end{center}
\end{figure}

A pioneering study of a parallel uncoupled array of bistable
oscillators has been performed with a general theory based on
linear response theory \cite{Neiman}, wherein the SNR gain is
below unity. Recently, Casado \textit{et al} reported that the SNR
gain is larger than unity for a mean-field coupled set of noisy
bistable subunits driven by subthreshold sinusoids
\cite{Casado2006}. However, each bistable subunit is subject to a
\textit{net} sinusoidal signal without input noise. The conditions
yielding a SNR gain exceeding unity have not been touched upon in
a parallel uncoupled array of bistable oscillators, in the
presence of either a subthreshold or suprathreshold sinusoid and
Gaussian white noise. In practice, an initially given noisy input
is often met, and a signal processor operating under this
condition, with the feature of the SNR gain exceeding unity, will
be of interest \cite{Chapeau2004,Chapeau2006}. The SNR gain has
been studied earlier in the less stringent condition of narrowband
noise \cite{Davenport}. In the present paper, we address the more
stringent condition of broadband white noise and the SNR gain
achievable by summing the array output, wherein extra array noise
can be tuned to maximize the array SNR gain. As the array size is
equal to or larger than two, the array SNR gain follows a SR-type
function of the array noise intensity. More interestingly, the
regions where the array SNR gain can exceed unity for a moderate
array size, are demonstrated numerically for both subthreshold and
suprathreshold sinusoids. Since the array SNR gain is amplified as
the array size increases from two to infinity, we can immediately
conclude that an infinite parallel array of bistable oscillators
has a global maximum array SNR gain for a fixed noisy sinusoid.
For an infinite parallel array, a tractable approach is proposed
using an array of two bistable oscillators, in view of the
functional limit of the autocovariance function \cite{Duan2006}.
We note that, for obtaining the maximum array SNR gain, the
control of this new class of array SR effect focuses on the
addition of array noise, rather than the input noise. This
approach can also overcome a difficult case confronted by the
conventional SR method of adding noise. When the initial input
noise intensity is beyond the optimal point corresponding to the
SR region of the nonlinear system, the addition of more noise will
only worsen the performance of system \cite{Xu}. Finally, the
optimization of the array SNR gain in an infinite array is touched
upon by tuning both an array parameter and array noise, and an
optimal array parameter is expected to obtain the global maximum
array SNR gain. These significant results indicate a series of
promising applications in array signal processing in the context
of array SR effects.
\section{\label{sec:II} The model and the array SNR gain}

The parallel uncoupled array of $N$ archetypal over-damped
bistable oscillators is considered as a model, as shown in
Fig.~\ref{fig:one}. Each bistable oscillator is subject to the
same signal-plus-noise mixture $s(t)+\xi(t)$, where
$s(t)=A\sin(2\pi t/T_s)$ is a deterministic sinusoid with period
$T_s$ and amplitude $A$, and $\xi(t)$ is zero-mean Gaussian white
noise, independent of $s(t)$, with autocorrelation $\left\langle
\xi(t)\xi(0)\right\rangle = D_{\xi}\delta (t)$ and noise intensity
$D_{\xi}$. At the same time, zero-mean Gaussian white noise
$\eta_i(t)$, together with and independent of $s(t)+\xi(t)$, is
applied to each element of the parallel array of size $N$. The $N$
array noise terms $\eta_i(t)$ are mutually independent and have
autocorrelation $\left\langle \eta_i(t)\eta_i(0)\right\rangle =
D_{\eta}\delta (t)$ with a same noise intensity $D_{\eta}$
\cite{Chapeau2006}. The internal state $x_i(t)$ of each dynamic
bistable oscillator is governed by
\begin{equation}
\tau_a\frac{dx_i(t)}{dt}=x_i(t)-\frac{x_i^3(t)}{X_b^2}+s(t)+\xi(t)+\eta_i(t),
\label{eq:one}
\end{equation}
for $i=1,2,\ldots,N$. Their outputs, as shown in
Fig.~\ref{fig:one}, are averaged and the response of the array is
given as
\begin{equation}
y(t)=\frac{1}{N}\sum_{i=1}^Nx_i(t). \label{eq:two}
\end{equation}
Here, the real tunable array parameters $\tau_a$ and $X_b$ are in
the dimensions of time and amplitude, respectively
\cite{Duan2004}. We now rescale the variables according to
\begin{equation}
 x_i(t)/X_b\rightarrow x_i(t),\; A/X_b\rightarrow A,
 \; t/\tau_a\rightarrow t, \; T_s/\tau_a\rightarrow T_s, \; D_{\xi}/(\tau_a X_b^2)\rightarrow D_{\xi},\; D_{\eta}/(\tau_a X_b^2)\rightarrow
 D_{\eta},
\label{eq:three}
\end{equation}
where each arrow points to a dimensionless variable.
Equation~(\ref{eq:one}) is then recast in dimensionless form as,
\begin{equation}
 \frac{dx_i(t)}{dt}=x_i(t)-x_i^3(t)+s(t)+\xi(t)+\eta_i(t).
\label{eq:four}
\end{equation}
Note that $s(t)$ is subthreshold if the dimensionless amplitude
$A<A_c=2/\sqrt{27}\approx 0.385$, otherwise it is suprathreshold
\cite{Peter2000,Duan2004}.

In general, the summed output response of arrays
$y(t)=(1/N)\sum_{i=1}^Nx_i(t)$ is a random signal. However, since
$s(t)$ is periodic, $y(t)$ will in general be a cyclostationary
random signal with the same period $T_s$ \cite{Chapeau1997}. A
generalized theory has been proposed for calculating the output
SNR \cite{Chapeau1997}. According to the theory in
\cite{Chapeau1997}, the summing response of arrays $y(t)$, at any
time $t$, can be expressed as the sum of its nonstationary mean
$E[y(t)]$ plus the statistical fluctuations $\tilde{y}(t)$ around
the mean $E[y(t)]$, as
\begin{equation}
y(t)=\widetilde{y}(t)+E[y(t)]. \label{eq:five}
\end{equation}
The nonstationary mean $E[y(t)]=(1/N)\sum_{i=1}^NE[x_i(t)]$ is a
deterministic periodic function of time $t$ with period $T_s$,
having the order $n$ Fourier coefficient
\begin{equation}
\overline{Y}_n=\left\langle E[y(t)]\exp(-\imath
2\pi\frac{n}{T_s})\right\rangle, \label{eq:six}
\end{equation}
where $\left\langle \cdots \right\rangle =(1/T_s)\int_0^{T_s}
\cdots dt$. For fixed $t$ and $\tau$, the expectation
$E[y(t)y(t+\tau)]$ is given by
\begin{equation}
E[y(t)y(t+\tau)]=E[\widetilde{y}(t)\widetilde{y}(t+\tau)]+E[y(t)]E[y(t+\tau)].
\label{eq:seven}
\end{equation}
Then, the stationary autocorrelation function $R_{\rm yy}(\tau)$
for $y(t)$ can be calculated by averaging $E[y(t)y(t+\tau)]$ over
the period $T_s$, as
\begin{eqnarray}
 R_{\rm yy}(\tau)&=& \left\langle E[y(t)y(t+\tau)]\right\rangle \nonumber \\
  &=& \left\langle E[\widetilde{y}(t)\widetilde{y}(t+\tau)]\right\rangle + \left\langle E[y(t)]E[y(t+\tau)]\right\rangle \nonumber \\
  &=& C_{\rm yy}(\tau)+ \left\langle E[y(t)]E[y(t+\tau)]\right\rangle, \label{eq:eight}
\end{eqnarray}
with the stationary autocovariance function $C_{\rm yy}(\tau)$ of
$y(t)$. The power spectral density $P_{\rm yy}(\nu)$ of $y(t)$ is
the Fourier transform of the autocorrelation function $R_{\rm
yy}(\tau)$
\begin{eqnarray}
  P_{\rm yy}(\nu) &=& \mathcal{F}[R_{\rm yy}(\tau)]=\int_{-\infty}^{+\infty}R_{\rm yy}(\tau)\exp(-\imath2 \pi\nu\tau)d\tau \nonumber \\
   &=& \mathcal{F}[C_{\rm yy}(\tau)]+
   \sum_{n=-\infty}^{+\infty}\overline{Y}_n\overline{Y}_n^*\delta(\nu-\frac{n}{T_s}). \label{eq:nine}
\end{eqnarray}
It is seen that the power spectral density $P_{\rm yy}(\nu)$ is
formed by spectral lines with magnitude $|\overline{Y}_n|^2$ at
coherent frequencies $n/T_s$, superposed to a broadband noise
background represented by the Fourier transform of $C_{\rm
yy}(\tau)$. Note that $E[\widetilde{y}(t)\widetilde{y}(t)]= \rm
{var}[y(t)]$ represents the nonstationary variance of $y(t)$,
which, after time averaging over a period $T_s$, leads to $C_{\rm
yy}(0)=\left\langle\rm {var}[y(t)]\right\rangle$, the stationary
variance of $y(t)$. The deterministic function $C_{\rm yy}(\tau)$
can thus be expressed as
\begin{equation}
C_{\rm yy}(\tau)=\left\langle\rm {var}[y(t)]\right\rangle h(\tau),
\label{eq:ten}
\end{equation}
where the correlation coefficient $h(\tau)$ is a deterministic
even function describing the normalized shape of $C_{\rm
yy}(\tau)$, having a Fourier transform
$\mathcal{F}[h(\tau)]=H(\nu)$. The power spectral density of
Eq.~(\ref{eq:nine}) can then be rewritten as
\begin{equation}
P_{\rm yy}(\nu)=\left\langle\rm {var}[y(t)]\right\rangle H(\nu)+
\sum_{n=-\infty}^{+\infty}\overline{Y}_n\overline{Y}_n^*\delta(\nu-\frac{n}{T_s}).
\label{eq:eleven}
\end{equation}
The output SNR is defined as the ratio of the power contained in
the output spectral line at the fundamental frequency $1/T_s$ and
the power contained in the noise background in a small frequency
bin $\Delta B$ around $1/T_s$, i.e.
\begin{equation}
R_{\rm out}(1/T_s)=\frac{|\overline{Y}_1|^2}{\left\langle\rm
{var}[y(t)]\right\rangle H(1/T_s)\Delta B}. \label{eq:twelve}
\end{equation}
In addition, the output noise is a Lorentz-like colored noise with
the correlation time $\tau_r$ defined by
\begin{equation}
h(|\tau|\geq \tau_r)\leq 0.05. \label{eq:thirteen}
\end{equation}

\begin{figure}
\begin{center}
\includegraphics[scale=0.55]{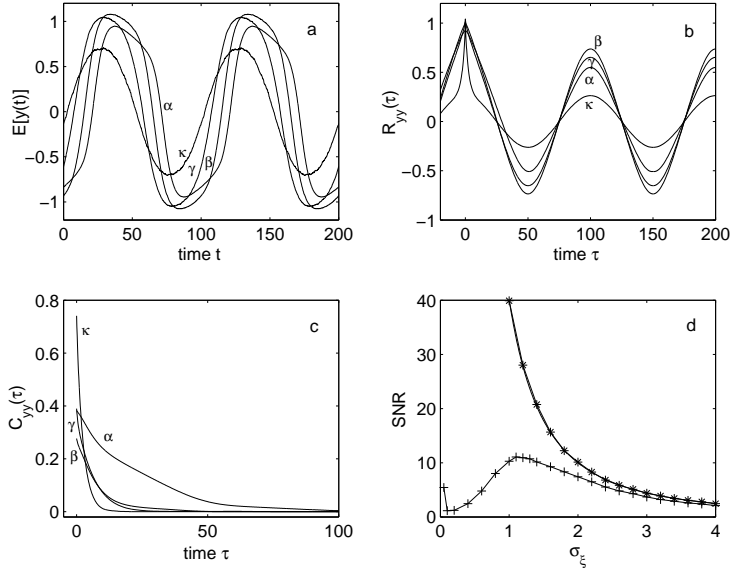}
\caption{Numerical output behaviors of a single bistable
oscillator for $A=0.4$ at four representative rms amplitudes
$\sigma_{\xi}$. (a) The nonstationary mean $E[y(t)]$. (b) The
stationary autocorrelation function $R_{\rm yy}(\tau)$. (c) The
stationary autocovariance functions $C_{\rm yy}(\tau)$. Here,
$\sigma_{\xi}=0.6$, $1.1$, $1.8$ and $3.0$ correspond to curves
\textbf{$\alpha$}, \textbf{$\beta$}, \textbf{$\gamma$} and
\textbf{$\kappa$}. (d) The theoretical input SNR (solid line) of
Eq.~(\ref{eq:fourteen}) as a function of $\sigma_{\xi}$. The
numerical input SNR $R_{\rm in}$ ($\ast$ but almost
indistinguishable) and output SNR $R_{\rm out}$ ($+$) are also
plotted. \label{fig:two}}
\end{center}
\end{figure}
In the same way, the periodic sinusoidal input $s(t)=A\sin(2\pi
t/T_s)$ has total power $A^2/2$ and power spectral density
$A^2[\delta(\nu+1/T_s)+\delta(\nu-1/T_s)]/4$ in the context of
bilateral power spectral density \cite{Chapeau1997}. Here, the
signal-plus-noise mixture of $s(t)+\xi(t)$ is initially given, and
the theoretical expression of input SNR can be computed as
\begin{eqnarray}
R_{\rm in}(1/T_s)=\frac{A^2/4}{D_{\xi}\Delta
B}=\frac{A^2/4}{\sigma^2_{\xi}\Delta t \Delta
B}.\label{eq:fourteen}
\end{eqnarray}
In the discrete-time implementation of the white noise, the
sampling time $\Delta t\ll T_s$ and $\tau_a$. The incoherent
statistical fluctuations in the input $s(t)+\xi(t)$, which
controls the continuous noise background in the power spectral
density, are measured by the variance
$\sigma_{\xi}^2=D_{\xi}/\Delta t$ \cite{Chapeau1997,Chapeau2006}.
Here, $\sigma_{\xi}$ is the rms amplitude of input noise $\xi(t)$.

Thus, the array SNR gain, viz. the ratio of the output SNR of
array to the input SNR for the coherent component at frequency
$1/T_s$, follows as
\begin{eqnarray}
G(1/T_s)=\frac{R_{\rm out}(1/T_s)}{R_{\rm
in}(1/T_s)}=\frac{|\overline{Y}_1|^2}{\langle \rm var
[y(t)]\rangle H(1/T_s)} \frac{\sigma^2_{\xi}\Delta t}{A^2/4}.
\label{eq:fifteen}
\end{eqnarray}
Equations~(\ref{eq:twelve})--(\ref{eq:fifteen}) can at best
provide a generic theory of evaluating SNR of dynamical systems
\cite{Chapeau1997}. If the array SNR gain exceeds unity, the
interactions of dynamic array of bistable oscillators and
controllable array noise provide a specific potentiality for array
signal processing. This possibility will be established in the
next sections.
\section{\label{sec:III} Numerical results of array SR and SNR gain}
We have carried out the simulation of parallel arrays of
Eq.~(\ref{eq:one}) and evaluated the array SNR gain of
Eq.~(\ref{eq:fifteen}), as shown in Appendix A, based on the
theoretical derivations contained in
\cite{Chapeau1997,Chapeau2006}. Here, we mainly present numerical
result as follows.
\begin{figure}
\begin{center}
\includegraphics[scale=0.55]{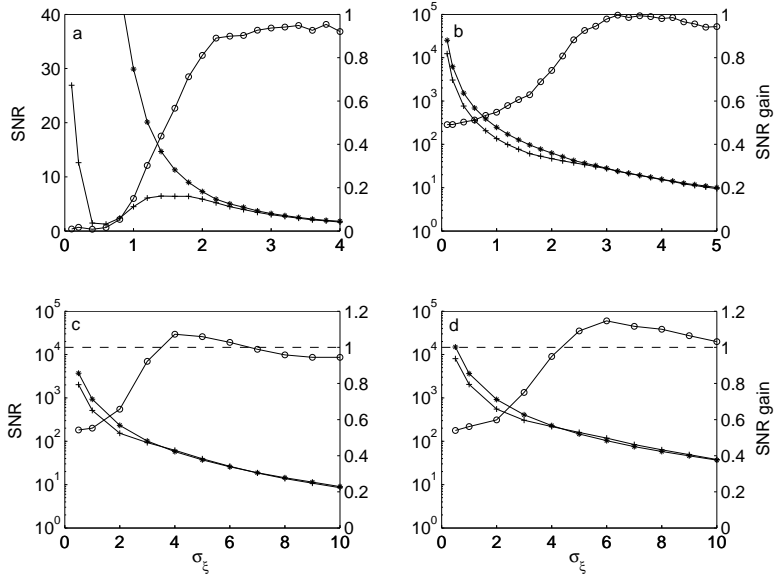}
\caption{Numerical input SNR $R_{\rm in}$ ($\ast$), output SNR
$R_{\rm out}$ ($+$) at the left axes and array SNR gain $G(1/T_s)$
($\circ$) at the right axes, as a function of $\sigma_{\xi}$, in a
single bistable oscillator for (a) $A=0.88A_c\approx 0.34$, (b)
$A=2.6A_c\approx 1.0$, (c) $A=5A_c\approx 1.925$ and (d)
$A=10A_c\approx 3.849$. Here $T_s=100$, $\Delta t \Delta
B=10^{-3}$, $\Delta t=T_s\times 10^{-3}$ and $K=10^{5}$.
\label{fig:three}} \end{center}
\end{figure}

\subsection{Improvement of the array SNR gain by noise for array size $N=1$}

If the array size $N=1$ and the response
$y(t)=(1/N)\sum_{i=1}^Nx_i(t)=x_1(t)$, this is the case of a
single bistable oscillator displaying the conventional SR or
residual SR phenomena \cite{Gammaitoni1998,Apostolico,Duan2004}.
In Figs.~\ref{fig:two}~(a)--(c), we show the evolutions of
$E[y(t)]$, $R_{\rm yy}(\tau)$ and $C_{\rm yy}(\tau)$,
respectively. The input is a sinusoidal signal with amplitude
$A=0.4$ and frequency $1/T_s=0.01$ mixed to the noise $\xi(t)$. As
the rms amplitude $\sigma_{\xi}$ increases, the periodic output
mean $E[y(t)]$ has a same frequency $1/T_s=0.01$, as shown in
Fig.~\ref{fig:two}~(a), and the largest amplitude of $E[y(t)]$
appears at the resonance region around $\sigma_{\xi}=1.1$. Plots
of the stationary autocovariance function $C_{\rm yy}(\tau)$, as
depicted in Fig.~\ref{fig:two}~(c), indicate that the correlation
time $\tau_{r}$ decreases as $\sigma_{\xi}$ increases, but the
stationary variance $C_{\rm yy}(0)=\langle\rm var [y(t)]\rangle$
presents a non-monotonic behavior. As $\sigma_{\xi}$ increases
from $0.6$ to $1.1$, $1.8$ and $3.0$, the correlation time
$\tau_r$ decreases from $65.1$ to $27.7$, $17.4$ and $7.5$,
whereas $C_{\rm yy}(0)$ equals to $0.383$, $0.277$, $0.390$ and
$0.741$, respectively. Thus, these nonlinear characteristics of
$E[y(t)]$ and $C_{\rm yy}(\tau)$ lead to the SR phenomenon of the
output SNR $R_{\rm out}$ versus the rms amplitude $\sigma_{\xi}$
in a single bistable oscillator, as illustrated in
Fig.~\ref{fig:two}~(d). The numerical input SNR $R_{\rm in}$ is
also plotted in Fig.~\ref{fig:two}~(d) and agrees well with the
theoretical one obtained by Eq.~(\ref{eq:fourteen}). Note that
this SR effect is residual SR introduced in
Ref.~\cite{Apostolico}, since the amplitude $A=0.4>A_c$ is
slightly suprathreshold. Similar results are presented in
Fig.~\ref{fig:three} for subthreshold amplitude ($A=0.34<A_c$) and
strong suprathreshold ones ($A=2.6A_c$, $5A_c$ and $10A_c$).
Clearly, the SR-type behaviors of $R_{\rm out}$ disappear for
strong suprathreshold amplitudes. These numerical results show
recurrence of the phenomena of conventional SR
\cite{Gammaitoni1998} and residual SR \cite{Apostolico}, and show
the validity of cyclostationary analysis presented in
Sec.\ref{sec:II} \cite{Chapeau1997}.

The SNR gain $G(1/T_s)$ is also depicted in Fig.~\ref{fig:three}
at the right axes. It is well known that the SNR gain $G(1/T_s)$
is below unity, so far as the sinusoidal amplitude $A$ is
subthreshold or slightly suprathreshold
\cite{Gammaitoni1998,Peter2000,Inchiosa1995,Jung1991,Dykman1998,Dykman1995},
as seen in Fig.~\ref{fig:two}~(d) and Fig.~\ref{fig:three}~(a).
However, as the amplitude $A$ increases to a more suprathreshold
value such as $A=2.6A_c$, $G(1/T_s)$ approaches unity very closely
at $\sigma_{\xi}=3.2$, as shown in Fig.~\ref{fig:three}~(b).
Interestingly enough, the possibility of $G(1/T_s)$ exceeding
unity exists for strong suprathreshold sinusoidal inputs ($A=5A_c$
or $10A_c$) at certain noise level regimes, as plotted in
Figs.~\ref{fig:three}~(c) and~(d). This result is consistent with
the work by H\"{a}nggi \textit{et al} \cite{Peter2000}.

These numerical results indicate that this theoretical framework
of cyclostationary signal processing in \cite{Chapeau1997} can
fully describe the SR phenomena in a single bistable oscillator,
and we shall now apply it to the SR effects in parallel uncoupled
arrays of bistable oscillators with a noisy sinusoidal input
$s(t)+\xi(t)$.

\subsection{Improvement by noise of the array SNR gain for array size $N\geq 2$}
\begin{figure}
\begin{center}
\includegraphics[scale=0.55]{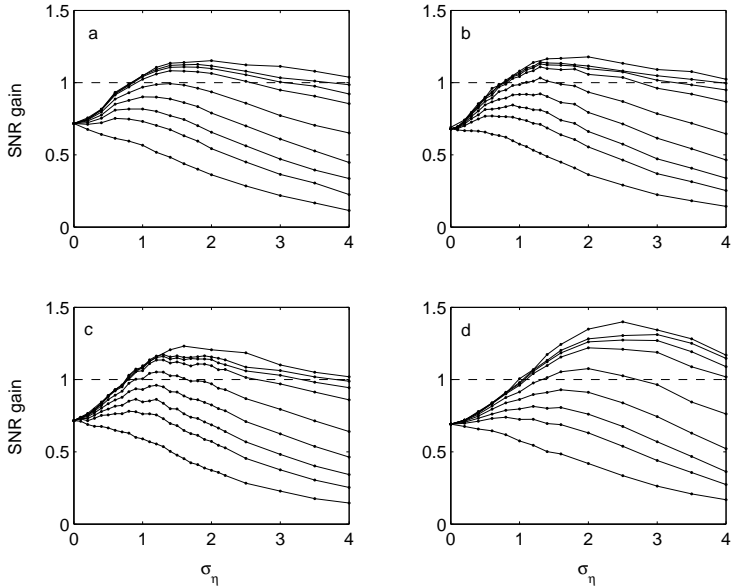}
\caption{Numerical array SNR gain as a function of the rms
amplitude $\sigma_{\eta}$ of array noise $\eta_{i}(t)$ for (a)
$A=0.34$, (b) $A=0.38$, (c) $A=0.4$ and (d) $A=1.0$. The array SNR
gain curves correspond to $N=1,2,3,5,10,30,60,120,\infty$ (from
the bottom up). The input noise rms amplitudes $\sigma_{\xi}=1.8$
for all amplitudes $A$, with given input SNRs $R_{\rm in}=8.92$,
$11.14$, $12.35$, and $77.16$, respectively. Here $T_s=100$,
$\Delta t \Delta B=10^{-3}$, $\Delta t=T_s\times 10^{-3}$ and
$K=10^{5}$. \label{fig:four}} \end{center}
\end{figure}
\begin{figure}
\begin{center}
\includegraphics[scale=0.35]{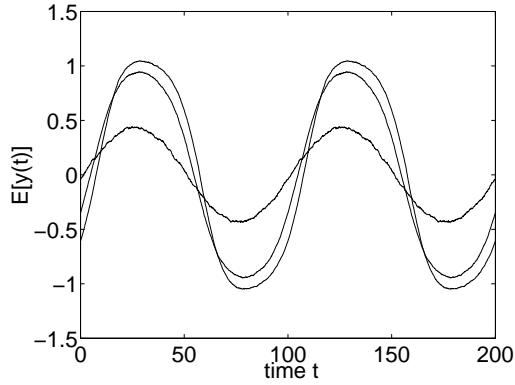}
\caption{Plots of nonstationary mean $E[y(t)]$ of $y(t)$ for
$A=0.4$ at $\sigma_{\eta}=0$, $1.3$ and $4.0$ (from the top down).
$E[y(t)]$ is a periodic function of time $t$ with period
$T_s=100$. Here, $t\in[0$, $2T_s[$ and $\Delta t=T_s\times
10^{-3}$. \label{fig:five}} \end{center}
\end{figure}
If the array size $N\geq 2$, this is the case of parallel arrays
of bistable oscillators displaying array SR phenomena.
Figure~\ref{fig:four} displays evolutions of the array SNR gain
$G(1/T_s)$ as a function of the rms amplitude $\sigma_{\eta}$ of
array noise $\eta_i(t)$, for both subthreshold ($A=0.34$ and
$0.38$) and suprathreshold inputs ($A=0.4$ and $1.0$). The input
noise rms amplitude is $\sigma_{\xi}=1.8$, resulting in the given
input SNRs $R_{\rm in}=8.92$, $11.14$, $12.35$, and $77.16$,
respectively. Then, due to array noise $\eta_i(t)$, the array SNR
gain $G(1/T_s)$ exhibits nonmonotonic behavior as a function of
$\sigma_{\eta}$ for $N\geq 2$. This collective phenomenon can be
termed as ``array SR'' \cite{Chapeau2006}, appearing for not only
suprathreshold inputs, as shown in
Figs.~\ref{fig:four}~(c)~and~(d), but also \textit{subthreshold}
signals, as presented in Figs.~\ref{fig:four}~(a)~and~(b). More
importantly, Fig.~\ref{fig:four} reveals that the region of the
array SNR gain $G(1/T_s)$ raising above unity, via increasing
$\sigma_{\eta}$, is possible for moderately large array size $N$.
Furthermore, as $A$ increases, $G(1/T_s)$ reaches a larger and
larger local maximal value for the same $N$. For instance,
$G(1/T_s)$ is about $1.1$ for $A=0.34$ and $N=120$, as shown in
Fig.~\ref{fig:four}~(a), whereas $G(1/T_s)$ is around $1.3$ for
$A=1.0$ and $N=120$, as seen in Fig.~\ref{fig:four}~(d).

The mechanism of conventional SR, as shown in
Figs.~\ref{fig:two}--\ref{fig:three}, exploits a combination of
the positive role of input noise $\xi(t)$ and the nonlinearity of
a single oscillator \cite{Gammaitoni1998,Peter2000}. Given a noisy
signal, the mechanism of array SR and the possibility of array SNR
gains above unity are clearly attributed to the added array noise
$\eta_i(t)$ interacting with the nonlinearity of the array
\cite{Chapeau2006}. Figure~\ref{fig:five} shows that nonstationary
means of $E[y(t)]$ are same for $N=1$, $2$, $\cdots$, $\infty$, at
fixed $\sigma_{\eta}$, since
\begin{equation}
E[y(t)]=E[\sum_{i=1}^Nx_i(t)/N]=\sum_{i=1}^NE[x_i(t)]/N=E[x_i(t)].
\label{eq:sixteen}
\end{equation}
However, we note that the amplitude of $E[y(t)]$ decreases as
$\sigma_{\eta}$ increases, as shown in Fig.~\ref{fig:five}. At
time $t$, we have
\begin{eqnarray}
R_{\rm yy}(\tau) &=& \left\langle E[y(t)y(t+\tau)]\right\rangle=
\left\langle E\Big[\frac{\sum_{i=1}^Nx_i(t)}{N}\cdot
\frac{\sum_{j=1}^Nx_j(t+\tau)}{N} \Big]\right\rangle \nonumber \\
&=& \left\langle
\frac{E[x_i(t)x_i(t+\tau)]}{N}+\frac{(N-1)E[x_i(t)x_j(t+\tau)]}{N}\right\rangle,
\label{eq:seventeen}
\end{eqnarray}
and
\begin{eqnarray}
C_{\rm yy}(\tau) &=&R_{\rm yy}(\tau) -\left\langle
E[y(t)]E[y(t+\tau)]\right\rangle=R_{\rm yy}(\tau)
-\left\langle \frac{E[\sum_{i=1}^Nx_i(t)]E[\sum_{j=1}^Nx_j(t+\tau)]}{N^2}\right\rangle \nonumber \\
&=& \left\langle
\frac{E[x_i(t)x_i(t+\tau)]}{N}+\frac{(N-1)E[x_i(t)x_j(t+\tau)]}{N}-E[x_i(t)]E[x_j(t+\tau)]\right\rangle,
\label{eq:eighteen}
\end{eqnarray}
for $i\neq j$ and $i,j=1,2,\ldots,N$. Note that
$E[x_i(t)]=E[x_j(t)]$.
\begin{figure}
\begin{center}
\includegraphics[scale=0.55]{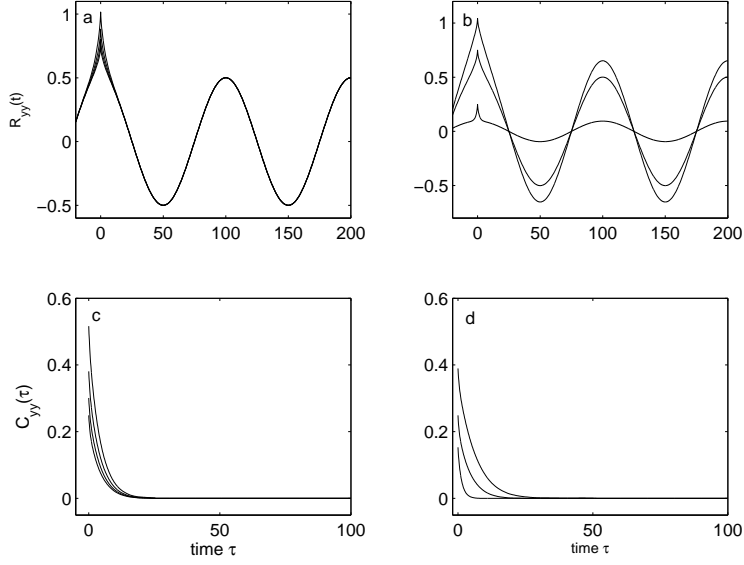}
\caption{Numerical behaviors of $R_{\rm yy}(\tau)$ and $C_{\rm
yy}(\tau)$. (a) $R_{\rm yy}(\tau)$ at $\sigma_{\eta}=1.3$ for
array sizes $N=1$, $2$, $5$ and $120$ (from the top down). (b)
$R_{\rm yy}(\tau)$ with array size $N=120$ as $\sigma_{\eta}$
varies from zero to $1.3$ and $4.0$ (from the top down). (c)
$C_{\rm yy}(\tau)$ at $\sigma_{\eta}=1.3$ for array sizes $N=1$,
$2$, $5$ and $120$ (from the top down). (d) $C_{\rm yy}(\tau)$
with array size $N=120$ as $\sigma_{\eta}$ changes from zero to
$1.3$ and $4.0$ (from the top down). Here, $A=0.4$, and other
parameters are the same as in Fig.~\ref{fig:four}.
\label{fig:six}} \end{center}
\end{figure}

Figures~\ref{fig:six}~(a) and~(c) show that, at
$\sigma_{\eta}=1.3$, $R_{\rm yy}(\tau)$ and $C_{\rm yy}(\tau)$
weaken as the array size $N$ increases. On the other hand, for a
fixed array size such as $N=120$, Figs.~\ref{fig:six}~(b) and~(d)
suggest that the output behaviors of $R_{\rm yy}(\tau)$ and
$C_{\rm yy}(\tau)$ also weaken as $\sigma_{\eta}$ increases from
$0$ to $1.3$ and $4.0$. Correspondingly, the stationary variance
$C_{\rm yy}(0)=0.39$, $0.25$ and $0.15$, and the correlation time
$\tau_r=17.4$, $12.1$ and $4.2$. An association of the time
evolutions of $E[y(t)]$ and $C_{\rm yy}(\tau)$ results in SR-type
curves of the array SNR gain $G(1/T_s)$ presented in
Fig.~\ref{fig:four}.

\begin{figure}
\begin{center}
\includegraphics[scale=0.3]{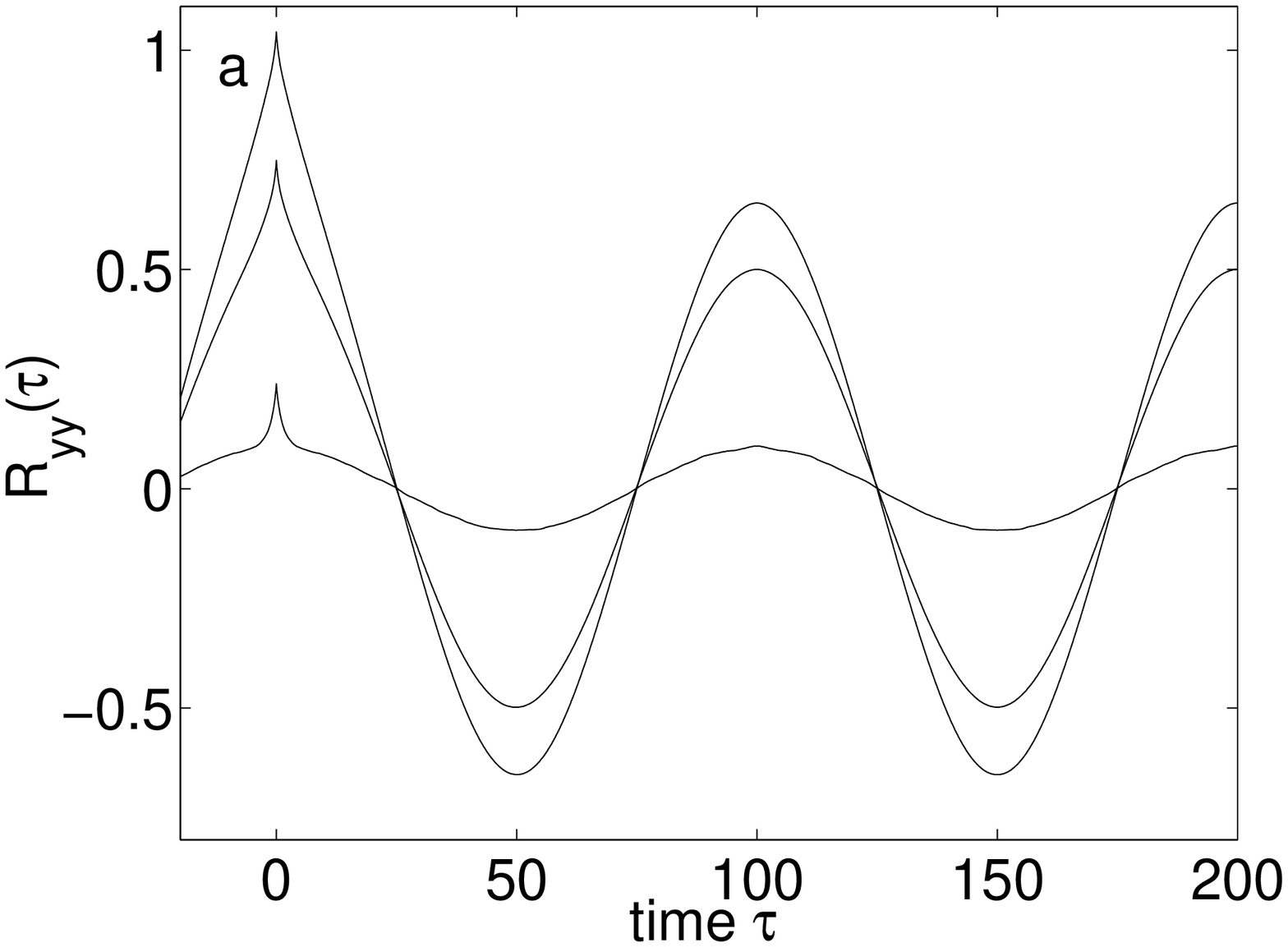}
\includegraphics[scale=0.3]{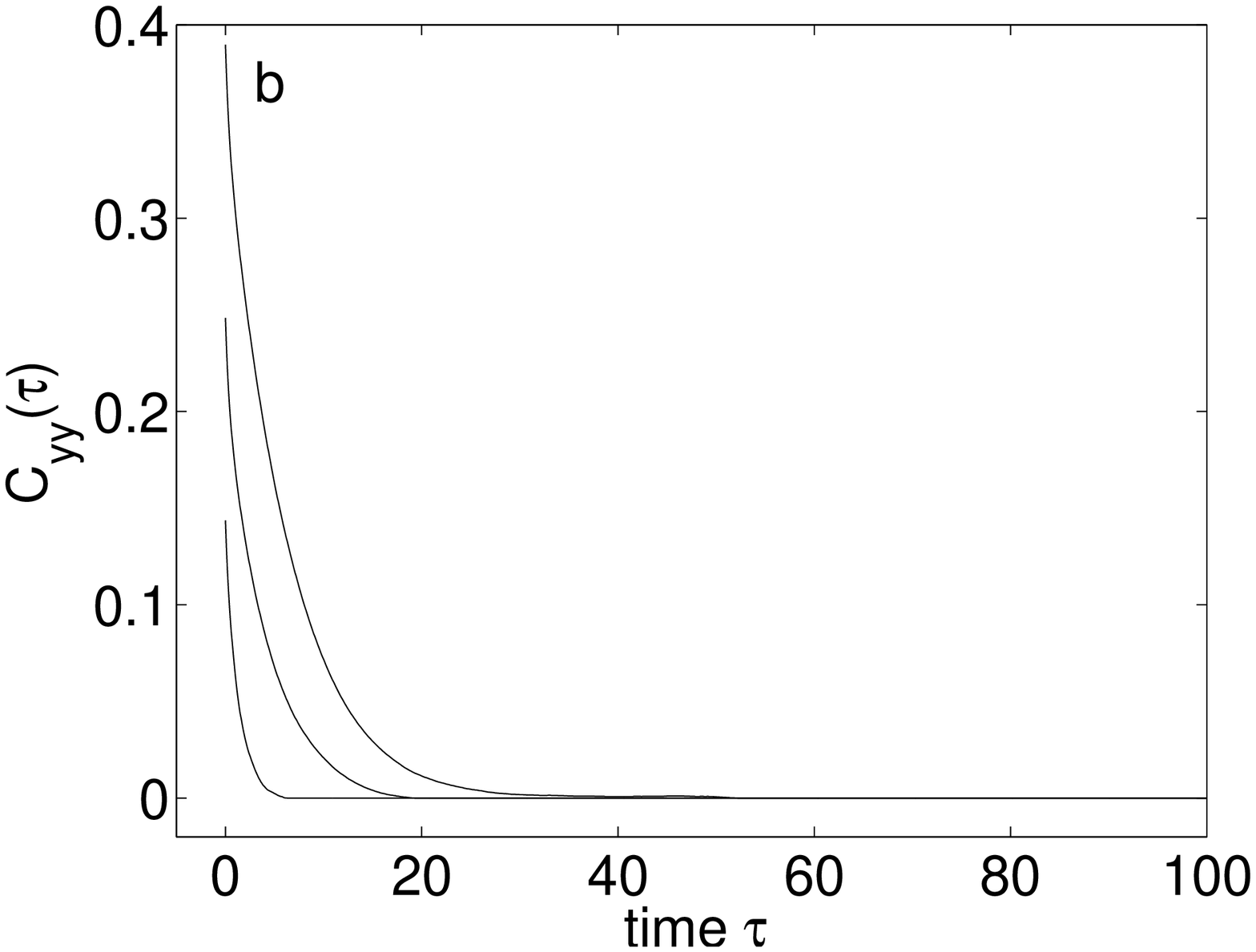}
\caption{Numerical output behaviors of arrays for array size
$N=\infty$. (a) $R_{\rm yy}(\tau)$ and (b) $C_{\rm yy}(\tau)$ as
functions of $\sigma_{\eta}=0$, $1.3$ and $4.0$ (from the top
down). Other parameters are the same as in Fig.~\ref{fig:four}.
\label{fig:seven}} \end{center}
\end{figure}

\subsection{Improvement by noise of the array SNR gain for array size $N=\infty$}
Figure~\ref{fig:four} shows that the array SNR gain $G(1/T_s)$ is
an increasing function of array size $N$. Thus, it is interesting
to know how much the maximal value of $G(1/T_s)$ reaches as array
size $N=\infty$. Form Eqs.~(\ref{eq:seventeen})
and~(\ref{eq:eighteen}), we have \cite{Duan2006}
\begin{eqnarray}
\lim_{N\rightarrow \infty} R_{\rm yy}(\tau) &=& \lim_{N\rightarrow
\infty}
\left\langle E[y(t)y(t+\tau)]\right\rangle \nonumber \\
&=& \lim_{N\rightarrow \infty} \left\langle
\frac{E[x_i(t)x_i(t+\tau)]+(N-1)E[x_i(t)x_j(t+\tau)]}{N}\right\rangle
\nonumber \\&=& \left\langle E[x_i(t)x_j(t+\tau)]\right\rangle
\nonumber \\&=& R_{x_i x_j}(\tau), \label{eq:nineteen}
\end{eqnarray}
and
\begin{eqnarray}
\lim_{N\rightarrow \infty} C_{\rm yy}(\tau) &=& \lim_{N\rightarrow
\infty} R_{\rm yy}(\tau) -\lim_{N\rightarrow \infty}
\left\langle E[y(t)]E[y(t+\tau)]\right\rangle \nonumber \\
&=& \lim_{N\rightarrow \infty} R_{\rm yy}(\tau) -
\lim_{N\rightarrow \infty} \left\langle
\frac{E[\sum_{i=1}^Nx_i(t)]E[\sum_{j=1}^Nx_j(t+\tau)]}{N^2}
\right\rangle \nonumber \\
&=& \left\langle E[x_i(t)x_j(t+\tau)]\right\rangle-\left\langle
E[x_i(t)]E[x_j(t+\tau)] \right\rangle \nonumber \\&=& C_{x_i
x_j}(\tau), \label{eq:twenty}
\end{eqnarray}
for $i\neq j$ and $i,j=1,2,\ldots,N$. Since the indices $i$ and
$j$ are different, but arbitrary in Eqs.~(\ref{eq:nineteen})
and~(\ref{eq:twenty}), we can adopt two bistable oscillators, each
embedded with independent noise, to evaluate the array SNR gain of
a parallel array with size $N=\infty$. The behaviors of $R_{\rm
yy}(\tau)$ and $C_{\rm yy}(\tau)$ are plotted in
Fig.~\ref{fig:seven} as the rms amplitude $\sigma_{\eta}$
increases from $0$, $1.3$ to $4.0$. The stationary variance
$C_{\rm yy}(0)=0.39$, $0.25$ and $0.15$, and the correlation time
$\tau_r=17.2$, $12.1$ and $4.2$, respectively. Furthermore, the
output SNR $R_{\rm out}$ of a parallel array of bistable
oscillators with infinite size $N=\infty$ is obtained from
Eq.~(\ref{eq:twelve}), and same for the array SNR gain $G(1/T_s)$
of Eq.~(\ref{eq:fifteen}). Numerical results of $G(1/T_s)$ are
also plotted in Fig.~\ref{fig:four} as $N=\infty$.

From Figs.~\ref{fig:four}--~\ref{fig:seven}, the mechanism of
array SR and the possibility of array SNR gain above unity can be
explained by the fact that independent array noise, on the one
hand, help the array response to reach its mean $E[y(t)]$, on the
other hand, counteract the negative role of input noise and
`whiten' the output statistical fluctuations $\widetilde y(t)$. In
other words, the stationary autocovariance function $C_{\rm
yy}(\tau)$ has a decreasing stationary variance $C_{\rm yy}(0)$
and correlation time $\tau_r$, as shown in
Figs.~\ref{fig:six}~and~\ref{fig:seven}.

\section{Optimization of the array SNR gain of an infinite array}
For a given input noisy signal and a fixed array size $N$, there
is a local maximal SNR gain, i.e.~the maximum value of $G(1/T_s)$
at the SR point of rms amplitude $\sigma_{\eta}$ of array noise,
as shown in Fig.~\ref{fig:four}. Clearly, this local maximal SNR
gain increases as array size $N$ increases, and arrives at its
global maximum $G_{\rm max}(1/T_s)$ as $N=\infty$. Note that
$G_{\rm max}(1/T_s)$ is obtained only via adding array noise
$\eta_i(t)$. It is interesting to know if $G_{\rm max}(1/T_s)$ can
be improved further by tuning both array noise $\eta_i(t)$ and the
array parameter $X_b$.

\begin{figure}
\begin{center}
\includegraphics[scale=0.5]{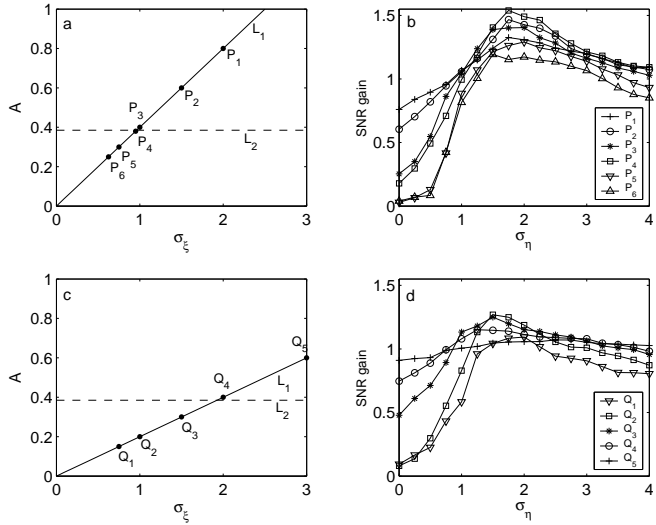}
\caption{(a) Plots of amplitude $A$ versus input noise rms
amplitude $\sigma_{\xi}$ (dimensionless variables). Line $L_1$ is
$A/\sigma_{\xi}=0.4$, and the corresponding input SNR $R_{\rm
in}=40$. Line $L_2$ of $A=2/\sqrt{27}\approx 0.385$ divides line
$L_1$ into subthreshold region (below $L_2$) and suprathreshold
section (over $L_2$). Points $P_i$ ($i=1,2,\cdots,6$) correspond
to $A=0.8$, $0.6$, $0.4$, $0.38$, $0.3$ and $0.25$, respectively.
(b) The global maximum SNR gain $G_{\rm max}(1/Ts)$, at fixed
input SNR $R_{\rm in}=40$, as a function of rms amplitude
$\sigma_{\eta}$ of array noise for points $P_i$ (different
amplitudes $A$). (c) Plots of $A$ versus $\sigma_{\xi}$. Line
$L_1$ is $A/\sigma_{\xi}=0.2$, and $R_{\rm in}=10$. Line $L_2$ is
$A=2/\sqrt{27}$. Points $Q_i$ ($i=1,2,\cdots,5$) correspond to
$A=0.15$, $0.2$, $0.3$, $0.4$ and $0.6$, respectively. (d) $G_{\rm
max}(1/Ts)$, at $R_{\rm in}=10$, as a function of $\sigma_{\eta}$
for points $Q_i$. Here, $T_s=100$, $\Delta B=1/T_s$ and $\Delta t
\Delta B=10^{-3}$. \label{fig:eight}} \end{center}
\end{figure}

In Eq.~(\ref{eq:three}), the signal amplitude $A/X_b\rightarrow A$
is dimensionless, and the discrete implementation of noise results
in the dimensionless rms amplitude of $\sigma_{\xi}/X_b\rightarrow
\sigma_{\xi}$ or $\sigma_{\eta}/X_b\rightarrow \sigma_{\eta}$
(where each arrow points to a dimensionless variable). The
dimensionless ratio of $A/\sigma_{\xi}$, as $\Delta t \Delta
B=10^{-3}$, determines the input SNR $R_{\rm in}$ of
Eq.~(\ref{eq:fourteen}). In Fig.~\ref{fig:eight}, we adopt two
given input SNRs $R_{\rm in}=40$ and $10$, this is,
$A/\sigma_{\xi}=0.4$ and $0.2$. When the array parameter $X_b$
varies, but $A/\sigma_{\xi}$ keeps, line $L_1$ comes into being,
and is divided into subthreshold region ($A<2/\sqrt{27}$) and
suprathreshold regime ($A>2/\sqrt{27}$) by line $L_2$ of
$A=2/\sqrt{27}$, as shown in Figs.~\ref{fig:eight}~(a) and~(c). We
select different points on line $L_1$, being located in
subthreshold region or suprathreshold region, for computing
$G_{\rm max}(1/T_s)$ via increasing $\sigma_{\eta}$, as
illustrated in Figs.~\ref{fig:eight}~(b) and~(d).
Figure~\ref{fig:eight}~(b) shows that, at the given input SNR
$R_{\rm in}=40$, the global maximum SNR gain $G_{\rm max}(1/T_s)$
increases from low amplitude $A=0.25$, i.e.~point $P_1$, reaches
its maximum around $\sigma_{\eta}=2.0$ for $A=0.38$, i.e.~point
$P_4$, then gradually decreases as the amplitude $A$ increases to
$0.8$ (point $P_1$). The same effect occurs for the given input
SNR $R_{\rm in}=10$, as shown in Fig.~\ref{fig:eight}~(d), and
$A=0.2$ (point $Q_2$) corresponds to the maximum $G_{\rm
max}(1/T_s)$ around $\sigma_{\eta}=1.5$. These results indicate
that, for a given input SNR, we can tune the array parameter $X_b$
to an optimal value, corresponding to an optimized global maximum
SNR gain.

However, we do not consider the other array parameter $\tau_a$,
which is associated with the time scale of temporal variables
\cite{Duan2004}. Then, the location of optimal array parameters
$X_b$ in subthreshold or suprathreshold regions, associated with
optimal $\sigma_{\eta}$, is pending. Immediately, an open problem,
optimizing the global maximum SNR gain $G_{\rm max}(1/T_s)$ via
tuning array parameters ($X_b$ and $\tau_a$) and adding array
noise (increasing $\sigma_{\eta}$), is very interesting but
time-consuming. This paper mainly focuses on the demonstration of
a situation of array signal processing where the parallel array of
dynamical systems can achieve a maximum SNR gain above unity via
the addition of array noise. Thus, the optimization of the maximum
SNR gain of infinite array is touched upon, and this interesting
open problem will be considered in future studies.

\section{Conclusions}
In the present work we concentrated on the SNR gain in parallel
uncoupled array of bistable oscillators. For a mixture of
sinusoidal signal and Gaussian white noise, we observe that the
array SNR gain does exceed unity for both subthreshold and
suprathreshold signals via the addition of mutually independent
array noise. This frequently confronted case of a given noisy
input and controllable fact of array noise make the above
observation interesting in array signal processing.

We also observe that, in the configuration of the present parallel
array, the array SNR gain displays a SR-type behavior for array
size larger than one, and increases as the array size rises for a
fixed input SNR. This SR-type effect of the array SNR gain,
i.e.~array SR, is distinct from other SR phenomena, in the view of
occurring for both subthreshold and suprathreshold signals via the
addition of array noise. The mechanism of array SR and the
possibility of array SNR gain above unity were schematically shown
by the nonstationary mean and the stationary autocovariance
function of array collective responses.

Since the global maximum SNR gain is always achieved by an
infinite parallel array at non-zero added array noise levels, we
propose a theoretical approximation of an infinite parallel array
as an array of two bistable oscillators, in view of the functional
limit of the autocovariance function. Combined with controllable
array noise, this nonlinear collective characteristic of parallel
dynamical arrays provides an efficient strategy for processing
periodic signals.

We argue that, for a given input SNR, tuning one array parameter
can optimize the global maximum SNR gain at an optimal array noise
intensity. However, another array parameter, associated with the
time scale of temporal variables is not involved. An open problem,
optimizing the global maximum SNR gain via tuning two array
parameters and array noise, is interesting and remains open for
future research.

\section{Acknowledgment}
Funding from the Australian Research Council (ARC) is gratefully
acknowledged. This work is also sponsored by ``Taishan Scholar''
CPSP, NSFC (No.~70571041), the SRF for ROCS, SEM and PhD PFME of
China (No.~20051065002).

\appendix
\section{Numerical method of computing power spectra of the collective
response of arrays} The corresponding measured power spectra of
the collective response $y(t)=(1/N)\sum_{i=1}^Nx_i(t)$ are
computed in a numerical iterated process in the following way that
is based on the theoretical derivations contained in
\cite{Chapeau1997,Chapeau2006}: The total evolution time of
Eq.~(\ref{eq:one}) is $(K+1)T_s$, while the first period of data
is discarded to skip the start-up transient
\cite{Chapeau,Peter2000}. In each period $T_s$, the time scale is
discretized with a sampling time $\Delta t\ll T_s$ such that
$T_s=L\Delta t$. The white noise is with a correlation duration
much smaller than $T_s$ and $\Delta t$. We choose a frequency bin
$\Delta B=1/T_s$, and we shall stick to $\Delta t \Delta
B=10^{-3}$, $T_s=100$, $L=1000$ and $K\geq 10^{5}$ for the rest of
the paper. In succession, we follow:

\noindent (\textbf{a}) The estimation of the mean $E[y(j\Delta
t)]$ is obtained over one period $[0$,$T_s[$, and the precise time
$j\Delta t$ of $E[y(j\Delta t)]$ ($j=0,1,\cdots,L-1$) shall be
tracked correctly in each periodic evolution of
Eq.~(\ref{eq:one}), i.e.~$[kT_s$, $(k+1)T_s[$ for
$k=1,2,\cdots,K$.

\noindent (\textbf{b}) For a fixed time of $\tau=i\Delta t$
($i=0,1,\cdots,\tau_{\rm max}/\Delta t$), the products $y(j\Delta
t)y(j\Delta t+i\Delta t)$ are calculated for
$j=1,2,\cdots,KT_s/\Delta t$. The estimation of the expectation
$E[y(j\Delta t)y(j\Delta t+i\Delta t)]$ is then performed. From
Eq.~(\ref{eq:eight}), the stationary autocorrelation function
$R_{\rm yy}(\tau)$ can be estimated over a time domain $\tau \in
[0$, $\tau_{\rm max}[$. Immediately, the stationary autocovariance
function $C_{\rm yy}(i\Delta t)$ of Eq.~(\ref{eq:eight}) at
$i=0,1,\cdots,\tau_{\rm max}/\Delta t$ can be deduced. Note the
time $\tau_{\rm max}$ is selected in such a way that at $\tau_{\rm
max}$, the stationary autocovariance function $C_{\rm yy}(i\Delta
t)$ in Eq.~(\ref{eq:eight}) has returned to zero. In practice, we
can select a quite small positive real number $\varepsilon$, such
as $\varepsilon=10^{-5}$. If $C_{\rm yy}(i\Delta t)/C_{\rm
yy}(0)\leq \varepsilon$, the above computation shall be ceased and
the index $i_{\rm end}$ is found, leading to $\tau_{\rm
max}=i_{\rm end}\Delta t$.

\noindent(\textbf{c}) Increase the total evolution time of
Eq.~(\ref{eq:one}) as $(K'+1)T_s$ ($K'>K$), and evaluate the mean
$E'[y(j\Delta t)]$ and the stationary autocovariance function
$C'_{\rm yy}(i\Delta t)$ again. If the differences between
$E'[y(j\Delta t)]$ and $E[y(j\Delta t)]$, $C'_{\rm yy}[i\Delta t]$
and $C_{\rm yy}(i\Delta t)$, converged within an allowable
tolerance, we go to the next step (\textbf{d}). If they do not
converge, the total evolution time of Eq.~(\ref{eq:one}) should be
increased to $(K''+1)T_s$ larger than $(K'+1)T_s$, until the
convergence is realized.

\noindent(\textbf{d}) With the converged mean $E[y(j\Delta t)]$
and stationary autocovariance function $C_{\rm yy}(i\Delta t)$,
the corresponding Fourier coefficient $\overline{Y}_1$ and the
power $\overline{\rm var [y(t)]}H(1/T_s)\Delta B$ of
Eq.~(\ref{eq:eight}) contained in the noise background around
$1/T_s$ can be numerically developed. The ratio of above numerical
values leads to the array SNR $R_{\rm out}$. The correlation time
$\tau_r=M\Delta t$ as $|h(M\Delta t)|=|C_{\rm yy}(M\Delta
t)/C_{\rm yy}(0)|\leq 0.05$. The numerical input SNR $R_{\rm in}$
can be also calculated by following steps
(\textbf{a})--(\textbf{d}), and compared with the theoretical
value of $R_{\rm in}$ of Eq.~(\ref{eq:thirteen}). The SNR gain
$G(1/T_s)$ will be finally figured out by Eq.~(\ref{eq:fifteen}).

\end{document}